\documentclass[
    ,final            % use final for the camera ready runs
  ]
  {aipproc}

\layoutstyle{6x9}

\begin{document}

\title{The Fate of Exoplanets and the Red Giant Rapid Rotator Connection}

  %97.10.Kc for stellar rotation, 97.20.Li for giant stars
              % \texttt{http://www.aip..org/pacs/index.html}>}

\author{Joleen K. Carlberg}{
  address={Department of Astronomy, University of Virginia, Charlottesville, VA 22904, USA}
}

\author{Steven R. Majewski}{
  address={Department of Astronomy, University of Virginia, Charlottesville, VA 22904, USA}
}

\author{Phil Arras}{
  address={Department of Astronomy, University of Virginia, Charlottesville, VA 22904, USA}
}
\author{Verne V. Smith}{
  address={National Optical Astronomy Observatory,  Tucson, AZ 85719, USA}
}

\author{Katia Cunha}{
  address={National Optical Astronomy Observatory,  Tucson, AZ 85719, USA}
  ,altaddress={Steward Observatory, 933 N. Cherry Ave.,  Tucson, AZ 85121} % additional visiting address
  ,altaddress={on leave from Observat\a'{o}rio Nacional, R. Gal. Jos\a'{e} Cristino, 77, 20921-400 S\~{a}o Crist\a'{o}v\~{a}o, Rio de Janeiro, RJ, Brazil} % additional visiting address
}

\author{Dmitry Bizyaev}{
  address={Apache Point Observatory, Sunspot, NM 88349, USA}
}

\begin{abstract}
 We have computed the fate of exoplanet companions around main sequence stars to explore the frequency of planet ingestion by their host stars during the red giant branch evolution.
Using published properties of exoplanetary systems  combined with stellar evolution models and Zahn's theory of tidal friction, we modeled the tidal decay of the planets' orbits as their host stars evolve. Most planets currently orbiting within 2~AU of their star are expected to be ingested by the end of their stars' red giant branch ascent. 
Our models confirm that many transiting planets are sufficiently close to their parent star that they will be accreted during the main sequence lifetime of the star. We also find that planet accretion may play an important role in explaining the mysterious red giant rapid rotators, although  appropriate planetary systems do not seem to be plentiful enough to account for all such rapid rotators.  We compare our modeled rapid rotators and surviving planetary systems to their real-life counterparts and discuss the implications of this work to the broader field of exoplanets.
 \end{abstract}

\maketitle

%%%%%%%%%%%%%%%%%%%%%%%%%%%%%%%%%%%%%%%%%%%%
%% MAINMATTER
%%%%%%%%%%%%%%%%%%%%%%%%%%%%%%%%%%%%%%%%%%%%

\section{Introduction}
As the number of known exoplanetary systems grows, we gain an ever more complete picture of the angular momentum reservoir stored in the exoplanetary orbits.   This reservoir can become important as the star evolves and begins to expand. At some point, many of the known exoplanets will be near enough to their host stars  for their gravity to raise tides on the star, which will distort the stellar shape and introduce a torque into the star-planet system. As long as the planets' orbital periods are shorter than the stellar rotation periods, the torque will act  in the sense that will ``spin-up'' the star. Tidal dissipation of energy in the convective envelopes of these now-red giant stars allows angular momentum to be transfered from the planetary orbit to the stars.  As the angular momentum is drained from the planetary orbit, the planet moves closer to the star, increasing the tidal distortion and accelerating the rate of the transfer.  As a result, the planet rapidly spirals into the star, dumping its angular momentum in the process.

The result of this planetary demise may help us understand the unusual class of rapidly rotating red giants.  Because stars should spin down as they evolve and expand, red giant stars are expected to have slow rotation speeds.  This expectation has been verified empirically by studies that find that most red giants are characterized by $v\sin i$~$\approx$~2~km~s$^{-1}$ \citep{gray81,gray82,demed96}.  A small fraction of red giants, around a few percent \citep[see, e.g.,][]{demed99,massarotti08a,carlberg10b}, deviate from this general rule.  They have $v\sin i$\ in excess of 10~km~s$^{-1}$ and sometimes significantly higher.  Many of  these stars have  no known stellar companions with which to interact, and planet accretion is a simple explanation that may account for these rapid rotators.  However, this explanation
raises  a number of questions  for which answers are needed to verify planet accretion as the underlying cause. 
 Does the number of rapid rotators predicted from modeling the future evolution of exoplanet systems match the number actually observed?  If not, what does this imply about the occurrence of planets around the progenitors of the red giant rapid rotators?  Can chemical abundances distinguish rapid rotators created by planet accretion from those created in some other way?
  
\section{Evolving  Main Sequence Stars with Planets}
\subsection{Tidal evolution model}
The present-day main sequence (MS) stars with known exoplanet companions can be used as a test progenitor population of red giant rapid rotators.
As an evolving star expands in radius, the separation between that star and  its planets shrinks.  As described in the introduction, once the planet is near enough to induce tides on the star then angular momentum can be exchanged between the stars and their planets. 
 The rate at which angular momentum is transfered depends on the rate at which energy is dissipated in the stellar envelope. These planet hosting stars are cool enough to have convective envelopes on the main sequence and will have convective outer atmospheres all throughout their red giant evolution.  Therefore, we use the model of turbulent dissipation described by \cite{zahn77}, where the friction timescale is estimated to be the eddy turnover timescale, $(MR^2/L)^{1/3}$, where $M$, $R$, and $L$ are the stellar mass, radius, and luminosity, respectively.
Using this model, the rate at which the planetary orbital separation decays is described by 
\begin{equation}
d\ln a\propto T_{\rm eff}^{4/3}M_{\rm env}^{2/3}M^{-1}q(1+q)(R/a)^8,
\label{eq1}
\end{equation}
where $T_{\rm eff}$ is the stellar effective temperature, $M_{\rm env}$ is the mass of the stellar convective envelope, $q$ is the ratio of the planet mass to the stellar mass, and $a$ is the orbital separation between the planet and star.
Notice that there is a  strong dependence on the relative size of the star compared to the star-planet separation; once the swelling star reaches some critical fraction of the orbital separation, the planet will rapidly spiral into the star.

The predicted evolution of the presently-known exoplanetary systems was accomplished by first matching each of the stars to a stellar evolution track using measured stellar mass and [Fe/H]  to match to the nearest model of $M$ and $Z$ in the \cite{giard00} grid of tracks.  The evolution of $a$ is then calculated from Equation \ref{eq1} at each time-step of the evolution models; the planet is assumed to have no effect on the evolution of the star.  Any angular momentum lost from a planet's orbit is added to the convective envelope of the star.   In this way, the rotational evolution of the star can be modeled from present time through to the red giant branch (RGB) tip.  Because of accelerated mass loss near the end of the RGB phase and the complexities of the helium flash, we do not model the evolution beyond the RGB tip.

To be selected for modeling, the planet-hosting (PH) stars must have a measured stellar mass and metallicity as well as measured planetary masses and semimajor axes.
In addition,  for the stars to be representative of present-day red giant stars, we only include those PH stars whose MS lifetimes are shorter than the age of the universe. This requirement essentially introduces a lower limit to the stellar mass, around 0.7~$M_{\odot}$, and has the largest effect in reducing the number of planetary systems eligible for consideration.  After applying these restrictions, 99  of the currently known exoplanet systems are available for our study---72 systems discovered with the radial velocity method and 27 discovered with the transit method.

\subsection{Results}
\begin{figure}
   \includegraphics[height=.48\textheight]{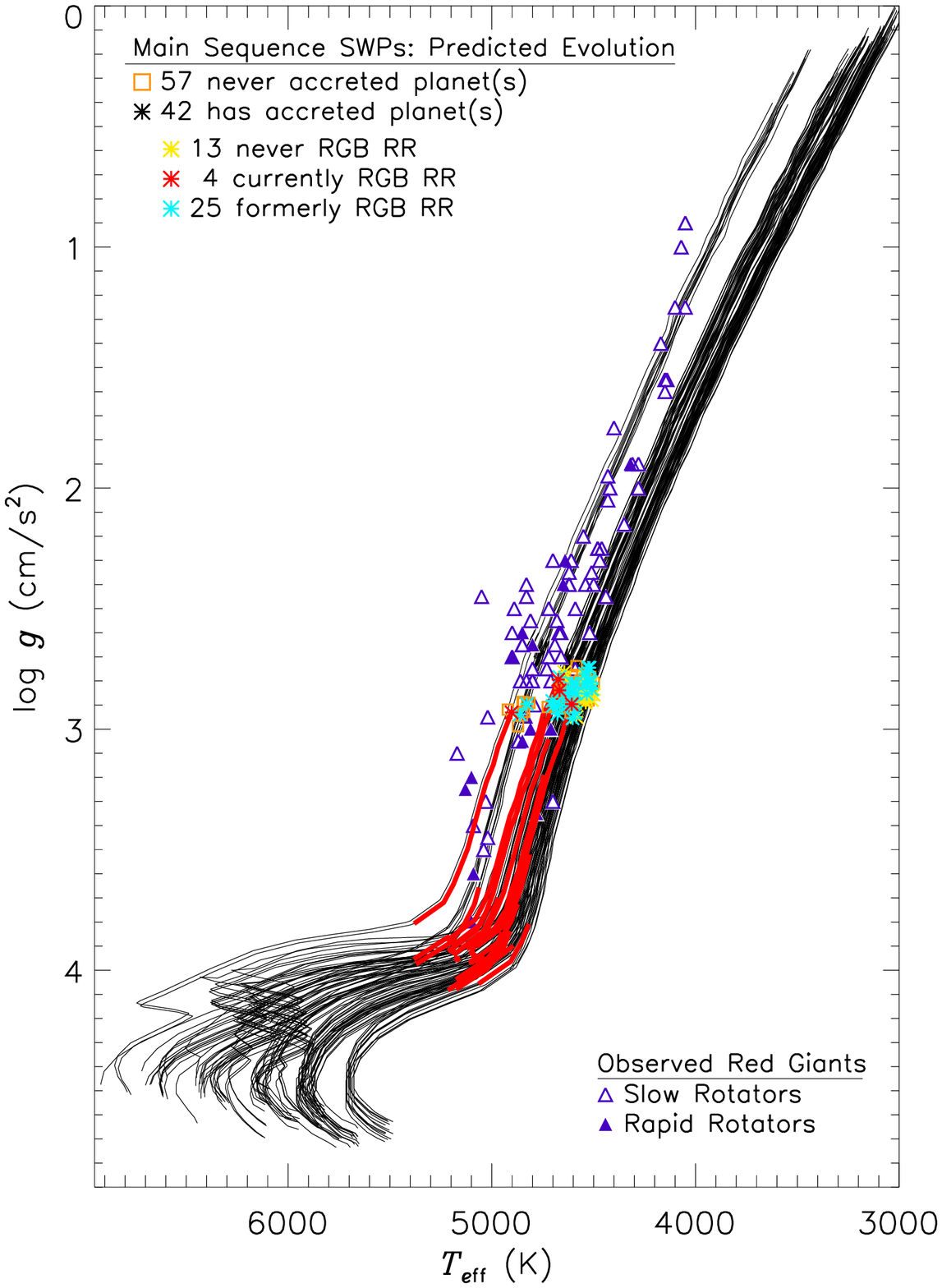}
   \includegraphics[height=.48\textheight]{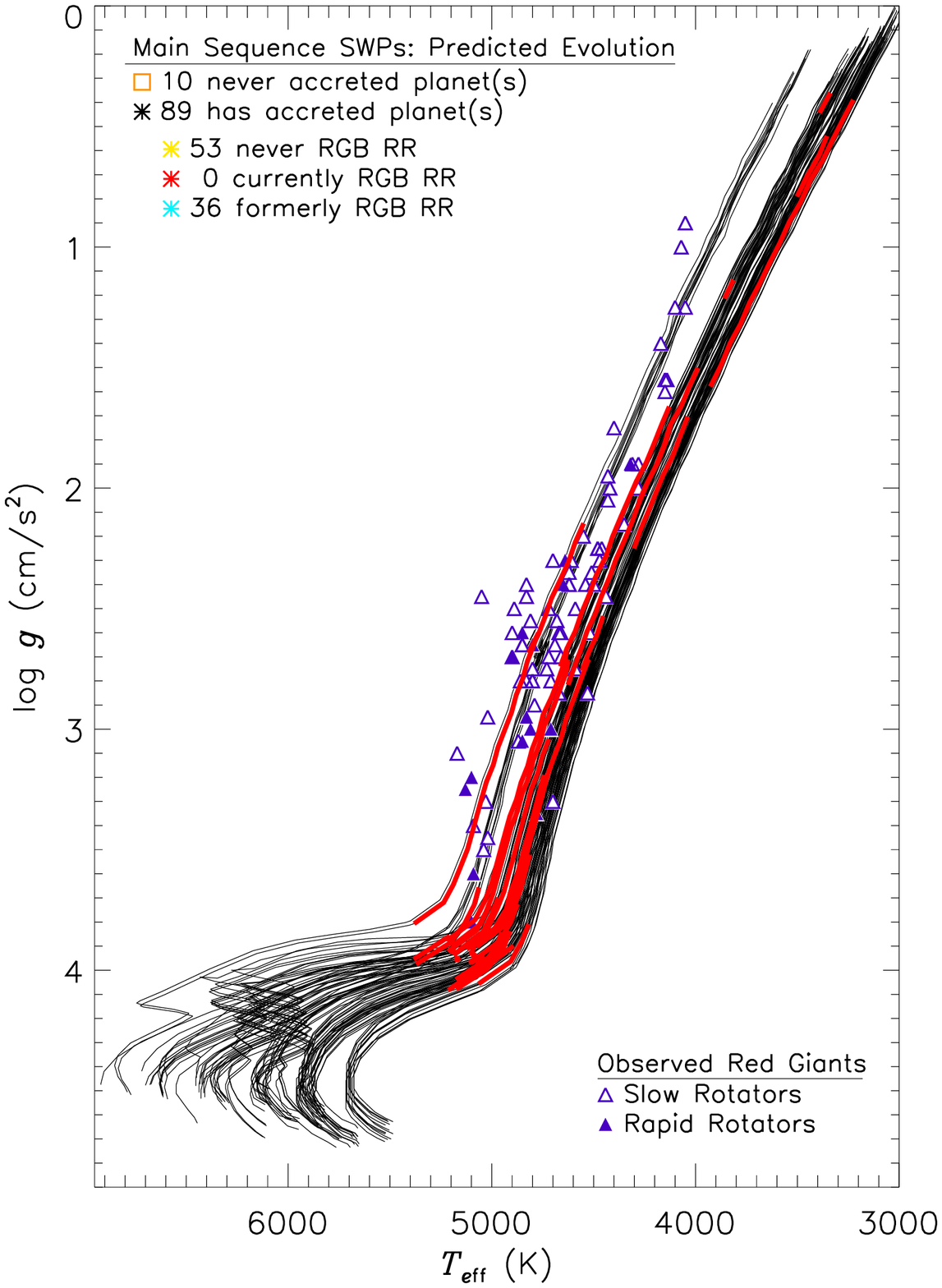}
  \caption{ A comparison of the temperature and surface gravity of the observed sample of red giants (purple triangles) compared to the simulated evolution of the MSs planet hosting stars for two ``snapshots'' in the evolution. The slow rotators are shown in open symbols and the filled symbols show the rapid rotators.    The black tracks  are the \cite{giard00} stellar evolution tracks for the exoplanet host stars. Because many of the PH stars have similar masses and metallicities and use the same evolution track, random offsets in $T_{\rm eff}$ and $\log g$ were applied to the tracks when plotting  for visibility. The squares and asterisks indicate whether the star has accreted a planet and the color of the asterisks further shows whether the star is/was/never was rapidly rotating.  The red strips on the evolution tracks show where the simulated PH stars were rapid rotators. 
  \label{fig:anim}}
\end{figure}

 Figure \ref{fig:anim} shows two snapshots in the evolution of the PH stars selected for this study.   The  $T_{\rm eff}$ and $\log g$ of the stellar evolution tracks for each PH star are shown, with small random offsets added to  each parameter for clarity. (Without these offsets, all stars matched to the same evolutionary track would overlap.)   The different symbols indicate whether the star has accreted any planets. The color of the symbols for stars that {\it have} accreted planets indicates whether that star is/was/never has been a RGB rapid rotator.  The red strips on the evolution tracks indicate that a star was a rapid rotator at that point in the evolution.  The purple triangles on the plot are observed giant stars, which will be discussed  in more detail later. %a later section.

The numbers in the second panel of Figure \ref{fig:anim} give the final statistics from our simulation, which show that of the 99 systems modeled,  89  stars will accrete one or more planets. However, only 36  stars will both accrete a planet and gain enough angular momentum on the red giant branch  to become a rapid rotator.   The stars that do become rapid rotators for at least part of the RGB evolution spend an average of 31\% of their RGB lifetimes as rapid rotators.  These numbers can be converted to a  prediction of how many rapid rotators are expected to be found in the RGB  population if they are created from planetary systems like the ones modeled here.   This expected rapid rotator fraction is the product of the fraction of stars with planets, the fraction of PH stars that become rapid rotators, and the fraction of the RGB lifetime  spent as a rapid rotator.  
If the fraction of PH stars is $\sim 5\%$ \citep{grether06}, then the expected fraction of rapid rotators is (5\%)(36\%)(31\%) = 0.56\%.  
\citep[A more detailed explanation of these models and results can be found in] []{carlberg09}.

This estimate of the expected fraction of rapid rotators is somewhat lower than the few-percent occurrence rate that is actually observed in the red giant population.   There are a number of reasons why this might be.  The first solution is simply the possibility that some of these rapid rotators, which were initially selected to be non-binary systems,  do in fact have an undetected binary companions with which they have interacted. 
Second, the known MS planet-hosting stars are not a perfect representation of the progenitors of present-day red giants.  The purple triangles in Figure \ref{fig:anim} come from our study to look for chemical evidence of planet accretion in known rapidly rotating giant stars  (Carlberg et al. in prep.).    
One of the first things noticeable about this sample of observed giant stars is that their temperatures and gravities do not entirely overlap with the evolution tracks of the exoplanet host stars.    This difference reflects the fact that the observed giants tend to be more massive and less metal-rich than the modeled PH stars.
Planet have been discovered around more massive stars, though the number known is still relatively small. Nevertheless, there is  growing evidence that these more massive PH stars have larger likelihoods of hosting planets than less massive stars \citep{johnson07a} and that that probability  is less dependent on metallicity \citep{pasquini07,hatzes08}.  If verified, these differences would translate to an increase in the expected fraction of rapid rotators in a PH star sample that is more representative of the observed red giant rapid rotators.

Even more interesting in Figure \ref{fig:anim} are the relative locations of the rapid rotators (filled symbols) and slow rotators (open symbols) compared to where  Figure \ref{fig:anim} indicates that  the PH stars are expected to be rapid rotators (i.e., the red strips).    The observed  rapid rotators tend to be less evolved and  lie in the $T_{\rm eff}$--$\log g$ plane where the density of the red strips (i.e., rapid rotator phase) on the evolutionary tracks tends to be highest. However, the most that can be said is that this behavior is {\it consistent} with the idea of planet accretion being responsible for the rapid rotators.

Of course,  proving that the correlation between predicted rapid rotators and RGB stars is a causal connection requires more substantial evidence,  which may be found in an another unusual property of red giant stars---lithium richness.
A well known consequence of stellar evolution is that lithium gets depleted in the stellar atmosphere.  The rate of this depletion depends on the depth of the convections layer (which dictates how deep the lithium is mixed, and thus the maximum temperature experienced by the lithium).  On the red giant branch, the phase of first dredge up causes significant lithium depletion.  Standard models \citep{iben67} predict depletion factors of around 50, % ($\Delta A$(Li)$=-1.7$), 
but most observed red giants show depletion factors far exceeding these theoretical predictions.   For example, \cite{lambert80} showed how the lithium abundance, $A$(Li)\footnote{$A$(Li)=$\log(N_{\rm Li}/N_{\rm H})+12$}, of a star with slightly super-solar mass and solar metallicity will evolve over the stellar lifetime.  $A$(Li) is initially near 3.3 dex.  By the end of the MS lifetime, $A$(Li) is reduced to around 1.1~dex, and at the tip of the red giant branch  it is further depleted to approximately $-0.8$~dex---a depletion of over four orders of magnitude!   Consequently, an accreted planet can significantly increase the lithium abundance in the stellar atmosphere if its lithium is evenly distributed in the stellar envelope.  
For a 1.25~$M_{\odot}$ RGB star with $A$(Li) $= -0.8$, an accreted Jupiter-mass planet can raise the stellar lithium abundance back up to 0.3 dex. %(triple check math!)

\begin{figure}
\includegraphics[height=.4\textheight]{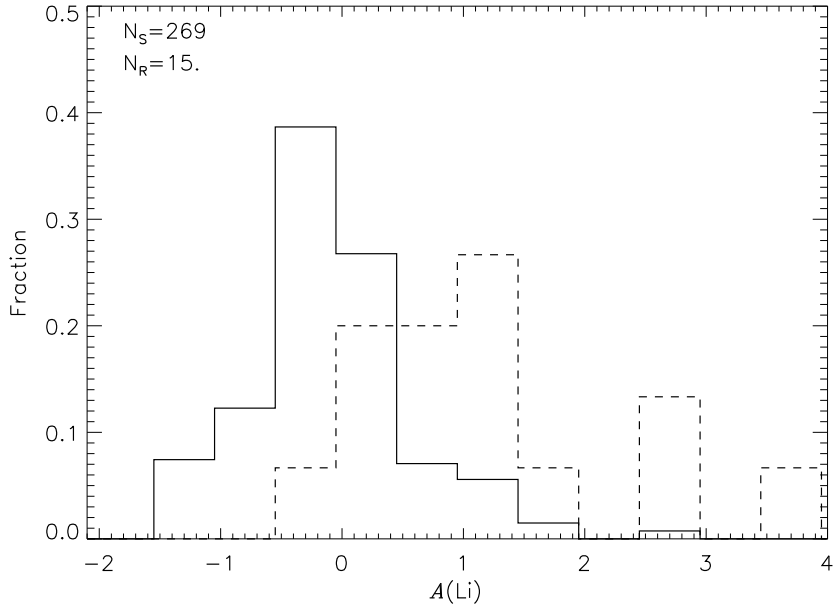}
  \caption{ Distribution of lithium abundances for the slow (solid line) and rapid (dashed line) rotators taken from the literature.  The 269 slow rotators come from \cite{demed00}, while the  fifteen rapid rotators were tabulated in \cite{drake02}.    \label{fig:histli}}
\end{figure}

The idea that both unusual lithium abundances and unusually fast rotation is attributable to an accreted planet was first put forth by \cite{alexander67} and has been brought up many times  to account for these atypical properties of giants \citep[e.g.,][]{wallerstein82,siess99,reddy02,drake02,carney03,denissenkov04,massarotti08a}.  
However, this explanation is still not universally accepted because stars often show one of these usual properties in the absence of the other. Consider Figure \ref{fig:histli}, which shows the lithium distribution of both rapid and slow rotators, taken from the literature.
The rapid rotators clearly tend to have higher lithium abundances than the slow rotators, but some slow rotators have high lithium, while some rapid rotators have low lithium.  A more compelling plot would be one that shows the difference between {\it expected} and {\it observed} $A$(Li), but  it is still not well understood what the expected values should be.  It should be remembered that the standard models actually predict less lithium dilution than what is typically observed.   Furthermore,  as \cite{drake02} rightly pointed out,  there can more than one mechanism at work to create lithium.

\section{Future Work}
There is much work to be done to understand not only the fate of exoplanetary systems but also the atypically rotating red giant stars. 
For the latter, looking for additional clues that rotation comes from accreted planets  may be found in a detailed comparison between these stars' abundance patterns and the normal rotators' abundances. However, 
a better understanding of normal stellar evolution may be needed  to ensure that any observed  abundance anomalies are truly anomalous for a given star.

To be sure, the success of looking for abundance changes from planet accretion is predicated on the assumption that once the planet enters common envelope phase, it is eventually evaporated and its material mixed throughout the stellar envelope.  How well is this common envelope phase understood?   Some detailed models of planet accretion have been done \citep{sandquist98,livio84,soker98} for a handful of specific cases, and an important conclusion drawn from these studies is that the outcomes vary significantly depending on the density gradient of the planet and on the stellar mass. Planets below a critical mass will evaporate while planets above some critical value actually accrete and grow. 

Finally, a better understanding of the fate of exoplanets may  also be obtained by a more thorough understanding of the distribution of planets around more massive stars. Radial velocity surveys of evolved PH stars have found a paucity of close orbiting planets \citep[e.g.,][]{johnson07b,johnson08,sato08,wright09}.  %from Villaver
\cite{villaver09} found that this paucity may be explained by tidal accretion; however, \cite{carlberg09} do not find this to be the case, especially if those PH giant stars are first ascent giants.  If the giants are not clearing out planets to the extent needed to explain the paucity seen in the observed radial velocity (RV)  surveys, then  it is likely that the lack of close-orbiting planets is related to the process of planet formation. The evolved stars probe much higher masses than the current sample of MS planet-hosting stars.  (Massive MS stars are poor RV targets because their hot atmospheres have few metal lines, and these stars tend to rotate rapidly; both make precise RV measurements difficult.)   Complementary planet searches for planets around more massive MS stars would determine whether close-orbiting planets do not form around massive stars, or if close-orbiting planets are all accreted early in their stars' post-MS evolution.

\begin{theacknowledgments}
  J.K.C. acknowledges support from the NASA Earth and Space Science Fellowship and the Virginia Space Grant Consortium.  Travel support was provided to J.K.C. from both the American Astronomical Society  International Travel Grant and from the conference.
 \end{theacknowledgments}

%%%%%%%%%%%%%%%%%%%%%%%%%%%%%%%%%%%%%%%%%%%%%%%%
%% The bibliography can be prepared using the BibTeX program or
%% manually.
%%
%% The code below assumes that BibTeX is used.  If the bibliography is
%% produced without BibTeX comment out the following lines and see the
%% aipguide.pdf for further information.
%%
%% For your convenience a manually coded example is appended
%% after the \end{document}
%%%%%%%%%%%%%%%%%%%%%%%%%%%%%%%%%%%%%%%%%%%%%%%%

%%%%%%%%%%%%%%%%%%%%%%%%%%%%%%%%%%%%%%%%%%%%%%%%
%% You may have to change the BibTeX style below, depending on your
%% setup or preferences.
%%
%%
%% For The AIP proceedings layouts use either
%%%%%%%%%%%%%%%%%%%%%%%%%%%%%%%%%%%%%%%%%%%%

\bibliographystyle{aipproc}   % if natbib is available
%\bibliographystyle{aipprocl} % if natbib is missing

%%%%%%%%%%%%%%%%%%%%%%%%%%%%%%%%%%%%%%%%%%%
%% You probably want to use your own bibtex database here
%%%%%%%%%%%%%%%%%%%%%%%%%%%%%%%%%%%%%%%%%%%

\end{document}